\documentstyle[a4]{article}

\begin{document}

\newcommand{\bib}{\bibitem}
\newcommand{\er}{\end{eqnarray}}
\newcommand{\br}{\begin{eqnarray}}
\newcommand{\be}{\begin{equation}}
\newcommand{\ee}{\end{equation}}
\newcommand{\epe}{\end{equation}}
\newcommand{\bea}{\begin{eqnarray}}
\newcommand{\eea}{\end{eqnarray}}
\newcommand{\ba}{\begin{eqnarray}}
\newcommand{\ea}{\end{eqnarray}}
\newcommand{\epa}{\end{eqnarray}}
\newcommand{\ar}{\rightarrow}
\newcommand{\dslash}{\partial\!\!\!/}
\newcommand{\aslash}{a\!\!\!/}
\newcommand{\eslash}{e\!\!\!/}
\newcommand{\bslash}{b\!\!\!/}
\newcommand{\vslash}{v\!\!\!/}
\newcommand{\rslash}{r\!\!\!/}
\newcommand{\cslash}{c\!\!\!/}
\newcommand{\fslash}{f\!\!\!/}
\newcommand{\Dslash}{D\!\!\!\!/}
\newcommand{\Aslash}{{\cal A}\!\!\!\!/}
\def\a{\alpha}
\def\b{\beta}
\def\r{\rho}
\def\D{\Delta}
\def\R{I\!\!R}
\def\l{\lambda}
\def\D{\Delta}
\def\d{\delta}
\def\T{\tilde{T}}
\def\k{\kappa}
\def\t{\tau}
\def\f{\phi}
\def\p{\psi}
\def\z{\zeta}
\def\ep{\epsilon}
\def\hx{\widehat{\xi}}
\def\na{\nabla}
\begin{center}

{\bf On the Group Structure of the Kalb-Ramond Gauge Symmetry.}

\vspace{1cm}

 M. Botta Cantcheff\footnote{e-mail: botta@cbpf.br}

\vspace{3mm}

 Centro Brasileiro de Pesquisas Fisicas (CBPF)

Departamento de Teoria de Campos e Particulas (DCP)

Rua Dr. Xavier Sigaud, 150 - Urca

22290-180 - Rio de Janeiro - RJ - Brasil.

\vspace{2mm}

Grupo de F\'{\i}sica Te\'orica Jose Leite Lopes, Petr\'opolis, RJ, Brazil.

\end{center}

\begin{abstract}

The transformation properties of a Kalb-Ramond field are those of
a gauge field; however, it is not clear which is the group
structure these transformations are associated with. The purpose
of this letter is to establish a basic framework in order to
clarify the group structure underneath the 2-form gauge potential.

\end{abstract}

\section{Introduction.}
The so-called (Abelian) Kalb-Ramond field \cite{kr}, $B_{\mu\nu}$,
is a two-form
 field which appears in the low energy limit of
String Theory and in several other frameworks in Particle Physics
\cite{aplic}; for instance,
 the most of the attempts to incorporate topological
 mass to the field theories in four dimensions take in account this object \cite{tm}.
  Its
dynamics is governed by an action which remains invariant under
transformations
 whose form are extremely similar to those of a one-form gauge field \cite{la}.
 The Kalb-Ramond (KR) field
transforms according with the following rule: \be B_{\mu\nu} \to
B_{\mu\nu} + \partial_{[\mu}\b_{\nu]}, \label{trKR} \ee where
$\b_\nu$ is a one form parameter. The question is, may this be
systematically generated from a group element? In other words, how
can we associate the parameter $\b_{\mu}$ to the manifold of some
gauge group? \cite{ultimo}. The first observation one can do is
that the set of parameters, namely
 $\Gamma\equiv\{\b_{\mu}\}$,
is not isomorphic to a group algebra in itself; however, we shall
show here how it encodes this information. The main purpose of
this letter is to shed some light to this question.

This work is organized as follows: in the Section 2, we deal with
the conditions to parametrize a group with an one-form parameter.
In Section 3; we considerably relax some apparently strong
restrictions, extending our analysis to multi-tensorial
parametrizations by considering a Clifford Algebra: similar
conditions are obtained and the structure of Section 2 is
recovered as a particular case. Concluding Remarks are collected
in Section 4; and finally, in an Appendix we explain the technical
details of our argument.

\section{A One-Form Group Parameter.}

Let us assume a four-dimensional Minkowski space-time
$(M,\eta_{\mu \nu})$ and some Lie group denoted by $G$, whose
associated algebra is ${\cal G}$; ${\tau}^{a}$ are the matrices
representing the generators of the group with $a= 1,\ldots ,
\mbox{dim}\:G$; $\tau_{abc}$ are the structure constants. Let us
take the gauge parameter to be an adjoint-one-form which can be
expanded as below: \be \b_\mu = \b^a_\mu \tau^a . \ee Consider
also a vector space ${\cal S}=\{\psi_I\}$ being a representation
for $G$, where $I,J$ denote the internal indices of this space
\footnote{Actually, $\psi_I$ is an $N$-component object, where $N$
is the the dimension of the group representation.}. Which is its
transformation law under a group element parametrized by the
object $\b^a_\mu$?. Let us take an infinitesimal transformation
$\ep^a_\mu \in \Gamma$. We shall get \be \psi'_I= g_{\ep} \psi = (
I_{IJ} + i \ep^a_\mu \tau^a \r^\mu_{IJ} + o^2(\ep) ) \psi_J
\label{1ord} \ee where $\{\r^\mu \}_{\mu=0}^3$ are four linearly
independent matrices which must transform as space-time vectors
under the Lorentz Group. For an arbitrary one form $\b_\mu$,
$\b_\mu \r^\mu$ must be a
 linear operator from ${\cal S}$ into itself. This remarkably implies
  that  ${\cal S}$ is some spinor space.


With the first order expression (\ref{1ord}), we may build up a
 group element corresponding to a non-infinitesimal parameter, $\b$;
 by considering $\b^a_\mu= N\, \ep^a_\mu$
for a large integer number $N$.
Thus, we have \be g_{\b} = \exp(i \b^a_\mu \tau^a \r^\mu). \ee For
the moment, we will not worry about the unitarity
 of these groups and their representations.

The key point in this construction arises from the analysis of the
group property \be g_{\b_1} g_{\b_2} = g_{\b_3} \label{compos}\, ,
\ee where $\b_{1,2,3} \in \Gamma$.

The question here is to identify under which conditions the parameters $\b_1$ and $\b_2$
 are such that $g_{\b_3}$ is also a group element.

Thus, one must have that $[\b_{1\,\mu} \r^\mu ; \b_{2\,\nu} \r^\nu ]$ is some linear
 combination of the group generators
$\tau^a$ (i.e, it remains in the algebra). This condition may be
written explicitly as \be \label{conm} [\b_{1} , \b_{2}]= \frac 12
\b^a_{1\,\mu} \b^b_{2\,\nu} \left([\tau^a , \tau^b] \{ \r^\mu ,
\r^\nu \}_{IJ} + \{\tau^a , \tau^b \} [ \r^\mu , \r^\nu ]_{IJ}
\right) = \b^a_{3} \tau^a , \ee where in principle, $\b^a_{3}$
also has to be a one-form.
 Since we cannot extend the
algebraic structure to be larger than that of the group
 $G$ and the
 space-time symmetries
(Lorentz group\footnote{Clearly, we must restrict our discussion
to space-time point-preserving transformations.}); in general, the
third term is not in the algebra, i.e, $[ \r^\mu , \r^\nu ]$ is
not a linear combination of the matrices $\r^\mu$, as well as the
anti-commutators
 $\{ \tau^a , \tau^b \}$
do not lie in ${\cal G}$ in general.

The single non-trivial algebraic structure we can count on,
compatible with the space-time symmetries,
 is the well-known Dirac's matrices algebra \footnote{Where the matrices $\gamma^\mu$ are
  linear maps on the Dirac spinor space in itself.}:
 \be \label{adirac} \{\gamma^\mu , \gamma^\nu \}_{IJ} = \eta^{\mu \nu} I_{IJ} ;\ee
 so for concreteness, all we can do is to identify the $\r^\mu$-matrices with the
  elements of this algebra
 and the fundamental spinorial representation, ${\cal S}$, with a Dirac spinor space.
 We shall carry out a more detailed
discussion about these points in the next section.

Thus, it became clear that the group property (\ref{compos}) {\it is not}
 satisfied for arbitrary
pairs of parameters $\b_1 , \b_2$.
 Therefore, we must determine which are the sub-families (sections of $\Gamma$)
of parameters, denoted by $P_{{\cal G}}$, such that it satisfy two basic requirements:

(i) For all pair $\b_1 , \b_2$ in the same $P_{{\cal G}}$, eq. (\ref{compos}) is satisfied
 for some $b_3 \in P_{{\cal G}}$ .

(ii) Each $P_{{\cal G}}$ is large enough to cover (parametrize) the group manifold.

Notice that due to relation (\ref{adirac}), it is natural to
introduce an
 additional scalar group parameter $\alpha$ associated with
 $I_{IJ}$; however, since the commutators $[\alpha, \beta] = \alpha^a \b^b_\mu
 [\tau^a , \tau^b]\r^\mu $ and $[\alpha , \alpha'] = \alpha^a \alpha'^b
 [\tau^a , \tau^b] $, remain in the algebra:
 no restrictions to this scalar parameters will arise. So, in order to
simplify our analysis, we stand for
  (\ref{compos}), (\ref{conm}) and the
following expressions involving the parameter $\b= \b^a_\mu \tau^a
\r^\mu_{IJ}$ {\it up to} the component associated to the identity.

For a better understanding of the conditions implied by the
requirements (i) and (ii),
 let us first consider
 the case of a Abelian group: $G \sim U(1)$.
Since no extra algebraic structure
 (of the type $[ \gamma^\mu , \gamma^\nu ]_{IJ} = c^{\mu \nu} I_{\,IJ} +
  c^{\mu \nu}_\r \gamma^\mu_{IJ}$, for certain tensors
   $c^{\mu \nu},c^{\mu \nu}_\r$\footnote{In particular, $I , \gamma^\mu ,
     [ \gamma^\mu , \gamma^\nu ]$ are
   linearly independents.})
can be required, the condition (\ref{conm}) may simply be
expressed as, \be \label{conmab} \b_{1\,\mu} \b_{2\,\nu} [
\gamma^\mu , \gamma^\nu ]_{IJ} = 0 . \ee
Then, we obtain a constraint on the parameters expressed by: \be
\b_{1\, [\mu} \b_{2\,\nu]} = 0 \label{paral} \ee
 This clearly implies that
$\b_{1\,\mu}$ and $\b_{2\,\nu}$ are {\it parallel}.

Of course, this condition is also sufficient to satisfy (i), (ii):
if $\b_{1\,\mu} \propto \b_{2\,\nu}$ the algebra closes and the
composition law (\ref{compos}) is satisfied. The other group
properties are automatically satisfied too. Then, we get that, for
each (fixed) direction in $\Lambda^1$ we choose, say $v_\mu$, we
have a group structure, and each straight line (a $dim
G$-dimensional plane in $\Gamma$) passing through the origin is
isomorphic to the group algebra ${\cal G}$ \footnote{Below, this
shall be confirmed for general Lie groups.}. Let us remark finally
that the choice of $v$ is completely arbitrary; so, this must not
be interpreted as a sort of gauge choice nor a breaking of the
Lorentz symmetry.
 For each direction $v$, one have the full structure of the group algebra.

For a more general Lie group, $G$, the second term of (\ref{conm})
must be vanish, therefore, the constraint on the group parameters
reads \footnote{Since this must not involve the generators, but
only the parameters.}: \be \b^{(a}_{1\, [\mu} \b^{b)}_{2\,\nu]} =
0 \, . \label{cons-cliff} \ee

One can give a proof (see appendix) that there are two families of
parameters which are solutions to these equations; the first one
reads \be\label{S0} \b^a_{1\,\mu} \propto \, \b^a_{2\,\mu} . \ee
Nevertheless, this does not satisfy a basic condition of the group
parametrization (requirement (ii)); this is a one-dimensional set
of parameters, while $dim G > 1$. The second family of solutions
is actually the appropriate one for us, \be \label{sol} \b^a_{\mu}
= \b^a \; v_{\mu} \,, \ee where $\b^a \in {\cal G}$, while
$v_{\mu} \in \Lambda_1$ has to be {\it fixed} for each group
parametrization.

Recalling the presence of an arbitrary term proportional to
$I_{IJ}$, the general form
 of an element of the algebra is
$ ( \alpha^a I + \b^a_{\mu}(= \b^a v_{\mu}) \gamma^\mu )$ for a
$v$ fixed. This algebra is: ${\bar{\cal G}}={\cal G} \oplus {\cal
G}$, the direct sum of two copies of ${\cal G}$ associated with
the matrices $I$ and $\vslash$ respectively.

So, our final result is that  \be\label{resultado} g(\alpha ,
\beta)= \exp{i\left( \alpha^a + (\b^a v_{\mu}) \gamma^\mu )\right)
\tau^a} \ee is a well-defined (generic) representative of a
element of the Lie group ${\bar G}$ which results from the
exponentiation of ${\bar{\cal G}}$. In the Abelian case this
simply reduces to ${\bar G} \sim G_{(I)}\circ G_{(\vslash)}$, the
{\it composition} of two copies of $G$ associated with $I$ and
$\vslash$ respectively.

For simplicity, it has been assumed up to now that only the five
matrices $\{I,\r^\mu\}$, appearing in (\ref{1ord}), can be used to
close the algebra (eq (\ref{conm})); but, since only a structure
compatible with the Lorentz symmetry may be considered, this
constitutes a strong restriction to the parameters as it has been
shown. However,
 one also count with the structure of the Clifford Algebra (CA)
 which closes by construction: so in principle,
 one may consider
additional independent gauge parameters (with other tensorial
ranks) apart from $\b_\mu$. We are going to see in the following
section that, even if we start off with such a more general
situation, one {\it necessarily} falls back to essentially the
same results, and the framework described above is actually not
too restrictive.

\section{Multi-tensorial Lie parameters via Clifford's Algebras
and fixed directions.}


Let us consider an ordered basis of the Clifford Algebra ${\cal
C}$: \be \{ \r^A \} \equiv \{ I , \gamma_5 , \gamma^{\mu} ,
\gamma_5 \gamma^{\mu}  , [ \gamma^\mu , \gamma^\nu ] \} ,\ee where
$A$ is a multi-index running over this multi-tensorial basis.
Thus we may represent infinitesimal group
transformations as

\be \psi'_I= g_{\ep} \psi = ( I + \ep^a_A \tau^a \r^A_{IJ} +
o^2(\ep) ) \psi_J ,\label{1ordcliff} \ee where $ \ep^a_A = ( \ep^a
, \ep'^a , \ep^a_\mu , \ep'^a_\mu , \ep^a_{\mu \nu} \}$
\footnote{These objects are also known in the literature as {\it
aggregates}}. The Clifford algebra satisfies a crucial property:
there exist a set of tensors, $C^{AB}_{\,\,C}$, such that
$(\r^A)_{IK} (\r^B)_{KJ} = C^{AB}_{\,\,C} (\r^C)_{IJ} $.


Notice once more that the component $\b^a$, associated with the
first element of the Clifford basis ($I_{IJ}$) does need not some
restriction, since $[ I , \r^B ]\equiv 0$; thus,  in order to
simplify our analysis as before, we stand for all the expressions
involving the multi-parameter ${\bf{\b}} = \b^a_A \tau^a
\r^A_{IJ}$ through this section, {\it up to} the component
associated to the identity.

Let us first consider an Abelian (one-dimensional) group like
$U(1)$:
 by construction, the commutator $[{\bf{\b}}_1 ; {\bf{\b}}_2]$ remains in ${\cal C}$.
  So, in principle,
one would have a $4\times4(=16)$-parametric representation of the
group. However, an important point has to be taken in account: if
no restrictions are imposed, these {\it are not} Abelian
representations. In fact, $\exp{\b_1} \exp{\b_2} \neq \exp{\b_2}
\exp{\b_1}$, unless precisely $[ \b_1 , \b_2]=0 $, which leads to
the extension of the conclusion of the previous section:
\be\label{sol0} \b_{(1,2)\, A} = \b_{(1,2)}\, V_A \, ,\ee for a
same (fixed) $V_{A} \in {\cal C}$. This is a central result in
this work: it states that, despite considering a full aggregate
structure for the parameters, we may only consider families of
parameters described by (\ref{sol0}), such as it was understood in
the first part of the article. In particular, if we wish to
involve only a one-form parameter as in the initial formulation of
the problem, one must to choose $V_{A}\equiv (0, 0, v_\mu , 0,
0)$. Recalling once more the presence of an arbitrary term
proportional to $I_{IJ}$, the more general form of the
multi-parameter involving a one-form is: \be \b= ( \alpha I +
\b_{\mu} \gamma^\mu ) \, ,\ee where the direction of $\b_\mu$ is
fixed, in coincidence with our previous result.

Notice that, if we consider an more general (non-Abelian) group
$G$, the above conclusion cannot be stated in the same way
because, by construction, the commutator \be \label{conmcliff}
2[\b_1 , \b_2] \equiv \b^a_{1\,A} \b^b_{2\, B} \left( [\tau^a ,
\tau^b] \{ \r^A , \r^B \} + \{\tau^a , \tau^b \} [ \r^A , \r^B ]
\right)_{IJ} \ee does not vanish. However, it may be extended also
to the non-Abelian case by observing that $\{\tau^a , \tau^b \}$
is representation-dependent and in general, it is out of the
algebra, thus the constraint must be imposed is similar to
(\ref{cons-cliff}): \be \label{cons-cliff2}\b^{(a}_{1\,[A}
\b^{b)}_{2\, B]}=0\; ,\ee whose solutions are also similar (see
Appendix): \be \label{sol2} \b^a_{A} = \b^a \; V_{A} \,, \ee where
$\b^a \in {\cal G}$, while $V_{A} \in {\cal C}$ has to be {\it
fixed} for each group parametrization. This is a $dim
G$-dimensional manifold of parameters such as it was understood in
the first part of the article. Again, we may choose $V_{A}\equiv
(0, 0, v_\mu , 0, 0)$ and due to the presence of an arbitrary
scalar gauge parameter, a generic group element writes as in
(\ref{resultado}).

\subsection{Remarks and Outline.}
The possible group structure associated with the KR symmetry was
determined and well-defined together with its spinorial
representations. We saw that, by considering the largest
multi-tensorial parameter space (Clifford Algebra), the result is
basically the same: the tensorial direction of the parameter is
{\it fixed}, and tensorially {\it separated} from the group
parameter (Eq.(\ref{sol})).
 This separability condition
for the Lie parameter is a bit unexpected, since it does not seem
to be a restriction arising from the transformation properties of
a KR-field. In a forthcoming article \cite{dkr}, we shall show
that, when one attempts to construct a covariant derivative, in
order to extract the KR-field from the connection,
 this condition desappears, i.e the total KR-gauge parameter is no longer separable.
  The reason is that the group structure
one must consider to define a covariant derivative is richer.

We conclude this letter by stressing that this construction is an
initial step, the fundamental starting point in order to construct
well-defined Abelian and non-Abelian gauge theories involving KR
fields. This open up the possibility of defining a gauge massive
model in four dimensions (even in the non-Abelian case), since
almost every theory with
 topological mass involves a rank-two gauge field.

In a future work we shall construct the covariant derivative and
extract from it, Kalb-Ramond two-form as a gauge field. Then, the
associated field strength is canonically defined and gauge
theories may normally be built up.

\vspace{0.5cm}

{\bf Aknowledgements}: The author is especially indebted to Alvaro
L. M. A. Nogueira for many invaluable discussions, relevant
comments and criticism. Prof. S. Alves D\'{\i}as and Patrick
Brockill are also acknowledged for discussions. Special thanks are
due also to J. A. Helayel-Neto for helpful and pertinent
corrections on the manuscript. Prof. Helayel-Neto pointed out
essential questions and motivations on an earlier work and
 observed the relation with the structure of aggregates .

\section{Appendix: On the possible families of group parameters.}

Here, we study the solutions to equation (\ref{cons-cliff2}) which
define a group parametrization. In the second section we found the
eq. (\ref{cons-cliff}) whose form is identical and the solutions
are of the
 same type, the single difference is that $\Lambda_1$ must be
 considered in behalf of the Clifford Algebra ${\cal C}$ analyzed here; thus,
 the statement proved here applies to the proposal of the Section 2.

  Let us denote by $P_{\cal G}$ a family of parameters,
${\bf{\b}}$, corresponding
 to a same group representation;
as has been argued before, if two parameters are in $P_{\cal G}$ then they are
 related by the equation (\ref{cons-cliff2}). Here we shall show
 that only class of solutions to this equation actually constitutes a group parametrization.

Contracting this equation with generic objects $X_a \in {\cal G}^{\star} $
 and $Y^A \in {\cal C}^{\star}$,
where ${\cal C}$ denotes the CA and ${}^{\star}$, the
corresponding {\it dual space},
 one obtains
the following relation between two elements of $P_{\cal G}$, $b_1$ and $b_2$:
\be
\label{a1}
\b^{a}_{2\,A} = \frac{b_2}{b_1} \, \b^{a}_{1\,A} + u^a\,U_A + w^a\,W_A \, ,
\ee
 for certain $u^a , w^a \in {\cal G}$ and $ U_A , W_A  \in {\cal C}$.
Here, we also have the complex numbers, $ b_{1,2} \equiv X_a Y^A
\b^{a}_{(1,2)\,A} $.

Next, we make two suppositions which shall define a particular class of solutions to
 (\ref{cons-cliff2}):

{\bf I}- $u^a , w^a$ are not proportional. Then, they may form,
together with
 ($dim ({\cal G}) - 2 $)-linearly independent
elements of ${\cal G}$, say $\{e_{i'}\}_{i'=1}^{dim ({\cal G}) - 2}$, a basis for it.

{\bf II}- $U_A , W_A  $ are not proportional. Then, they may form,
together with $(dim ({\cal C}) - 2 (=14)$)-linearly independent
elements of ${\cal C}$, say $\{C_{i A}\}_{i=1}^{14}$, a basis for
it.

Clearly, we may write \be \label{a2} \b^{a}_{1\,A} =
\Sigma_{i=1}^{14} a_i^a C_{i A}  +  a_U^a U_A + a_W^a W_A .\ee
Substituting this into the equation (\ref{cons-cliff2}), and
 using the identity $\b^{(a}_{1\,[A} \b^{b)}_{1\,B]} \equiv 0 $, we get the equation
\be \label{a3}
 \Sigma_{i=1}^{14} [ a_i^{(a}u^{b)} C_{i [A} U_{B]}  + a_i^{(a}w^{b)} C_{i [A} W_{B]}]
  + [ a_W^{(a}u^{b)} -
a_U^{(a}w^{b)}]W_{[A} U_{B]}  =0 ,\ee which, due to the linear
independence implies $a_i^{(a}u^{b)} =0$, $a_i^{(a}w^{b)} =0$ and
\be \label{a4} a_W^{(a}u^{b)} - a_U^{(a}w^{b)}=0 .\ee Contracting
the first two equations with an arbitrary element of ${\cal G}$,
one obtains that $a_i^{a}$ must be proportional to both $u^a ,
w^a$; since they are not parallel by hypothesis, we conclude
$a_i^{a}=0$.

So, we shall finally solve only the equation (\ref{a4}). Due to
the assumption {\bf I}, one may write \be \label{a5} a^{a}_{U} =
\Sigma_{i=1}^{dim {\cal G} - 2 } a_U^{i'} e_{i'}^a  +  a_U^{u} u^a
+ a_U^{w} w^a \ee and \be \label{a6} a^{a}_{W} = \Sigma_{i=1}^{dim
{\cal G} - 2 } a_W^{i'} e_{i'}^a  +  a_W^{u} u^a + a_W^{w} w^a \ee
where $a_U^{i'} , a_W^{i'} , a_U^{u} , a_U^{w} , a_W^{u} ,
a_W^{w}$ are complex numbers. Putting these expressions back into
eq. (\ref{a4}) we get \be \label{a7} \Sigma_{i=1}^{dim {\cal G} -
2 } [ a_U^{i'} e_{i'}^{(a} w^{b)} - a_W^{i'} e_{i'}^{(a} u^{b)} ]
+ [a_U^{u} - a_U^{w}] u^{(a}w^{b)} + a_U^{w} w^{(a}w^{b)} +
a_W^{u} u^{(a}u^{b)}=0 . \ee
 Due to {\bf I}, we conclude that $a_U^{i'}=0 , a_W^{i'}=0  , a_U^{w}=0  , a_W^{u}=0  $ and
$a\equiv a_U^{u} = a_U^{w}$. Then, our multi-parameter initial
must have the form \be \label{a8} \b^{a}_{1\,A} = a [ u^a\,U_A +
w^a\,W_A ]; \ee which, by virtual of eq. (\ref{a1}), implies \be
\label{a9} \b^{a}_{2\,A} \propto \b^{a}_{1\,A} \ee However, as it
has already been argued, this class of solutions do not constitute
a group parametrization because it is a one-parametric family of
parameters; therefore, it is not isomorphic to ${\cal G}$ (unless
it is a one-dimensional group, which was excluded from this
discussion).

So, we only can consider some of the hypotheses {\bf I}, {\bf II}
to be false; in both cases, eq. (\ref{a1}) may be rewritten as \be
\label{a10} \b^{a}_{2\,A} = \frac{b_2}{b_1} \, \b^{a}_{1\,A} +
v^a\,V_A \; , \ee
 for certain $v^a \in {\cal G}$ and $ V_A \in {\cal C}$. Then, we may write
\be \label{a11} \b^{a}_{1\,A} = \Sigma_{i=1}^{15} a_i^a C_{i A}  +
\b_1^a V_A \; ,\ee where $\{ \{ C_{i A}\}_{i=1}^{dim ({\cal C}) -1
(=15)} \; ; V_A \}$ constitutes a basis for ${\cal C}$. Thus,
 using this in the equation (\ref{cons-cliff2}), one gets
\be
\label{a12}
 \Sigma_{i=1}^{15}  a_i^{(a}v^{b)} C_{i [A} V_{B]}  =0 .
\ee
Therefore, \be \label{a13} a_i^{(a}v^{b)}=0\ee. Contracting this equation
 with an arbitrary element of ${\cal G}$:
$a_i^{a}=a_i \, v^{b}$; plugging this back into (\ref{a13}), we
get $a_i^{a}=0$.
 Then, the complete solution
results as below:\be \b^{a}_{1\,A} = \b_1^a V_A \; . \ee From
(\ref{cons-cliff2}), $\b^{a}_{2\,A} = \b_2^a V_A $, which shows
that all parameter in this family writes as $\b^{a}_{A} = \b^a V_A
$, where $V_A$ is fixed. This is a $dim {\cal G}$-parametric
family of parameters, as expected.

This completes the argument.



\begin{thebibliography}{99}
\bibitem{kr} M.~Kalb and P.~Ramond,
Phys.\ Rev.\ D {\bf 9}, 2273 (1974).

D.~Z.~Freedman, CALT-68-624. K.~Seo, M.~Okawa and A.~Sugamoto,
Phys.\ Rev.\ D {\bf 19}, 3744 (1979).

D.~Z.~Freedman and P.~K.~Townsend,
Nucl.\ Phys.\ B {\bf 177}, 282 (1981).



\bibitem{aplic}
A.~S.~Schwarz,
Lett.\ Math.\ Phys.\  {\bf 2}, 247 (1978).

A.~S.~Schwarz,
Commun.\ Math.\ Phys.\  {\bf 67}, 1 (1979).

E.~Witten,
Commun.\ Math.\ Phys.\  {\bf 117}, 353 (1988).

D.~Birmingham, M.~Blau, M.~Rakowski and G.~Thompson,
Phys.\ Rept.\  {\bf 209}, 129 (1991).



\bibitem{tm}
E.~Cremmer and J.~Scherk,
Nucl.\ Phys.\ B {\bf 72}, 117 (1974).

A.~Aurilia and Y.~Takahashi,
Prog.\ Theor.\ Phys.\  {\bf 66}, 693 (1981).

T.~R.~Govindarajan,
J.\ Phys.\ G {\bf 8}, L17 (1982).

T.~J.~Allen, M.~J.~Bowick and A.~Lahiri,
Phys.\ Lett.\ B {\bf 237}, 47 (1990).

T.~J.~Allen, M.~J.~Bowick and A.~Lahiri,
Mod.\ Phys.\ Lett.\ A {\bf 6}, 559 (1991).




\bibitem{la} A. Lahiri, Mod.Phys.Lett.A17 (2002) 1643;
J.Phys.A35 (2002) 8779, and references therein.


\bibitem{ultimo} C. Hofman, {\it Nonabelian 2 Forms},
 e-Print Archive: hep-th/0207017.

\bibitem{dob} M. Botta Cantcheff,
Phys. Lett. B 533 (2002) 126 and Eur. Phys. Jour. C6 vol.4
(2002)1-14 (e-Print Archive: hep-th/0107123).


\bibitem{dkr} M. Botta Cantcheff, {\it The Kalb-Ramond Field as a Connection
on a Flat Space Time.}, work in progress.


\end{thebibliography}
\end{document}